\documentclass[twocolumn,showpacs,preprintnumbers,amsmath,amssymb,superscriptaddress,prl,floatfix]{revtex4}

\usepackage{graphicx}% Include figure files
\usepackage{dcolumn}% Align table columns on decimal point
\usepackage{bm}% bold math
\usepackage{longtable}

\begin{document}

\setlength{\parskip}{0 pt}

\preprint{APS/123-QED}

\bibliographystyle{unsrt}

\title{Droplet-like Fermi surfaces in the anti-ferromagnetic phase of EuFe$_2$As$_2$, an Fe-pnictide superconductor parent compound}

\author{S. de Jong}
\email{sdejong@science.uva.nl}
\affiliation{Van der Waals-Zeeman Institute, University of Amsterdam, NL-1018XE
Amsterdam, The Netherlands}

\author{E. van Heumen}
\affiliation{Van der Waals-Zeeman Institute, University of Amsterdam, NL-1018XE
Amsterdam, The Netherlands}

\author{S. Thirupathaiah}
\affiliation{Helmholtz Zentrum Berlin, Albert-Einstein-Strasse 15, 12489 Berlin, Germany}

\author{R. Huisman}
\affiliation{Van der Waals-Zeeman Institute, University of Amsterdam, NL-1018XE Amsterdam, The Netherlands}

\author{F. Massee}
\affiliation{Van der Waals-Zeeman Institute, University of Amsterdam, NL-1018XE Amsterdam, The Netherlands}

\author{J. B. Goedkoop}
\affiliation{Van der Waals-Zeeman Institute, University of Amsterdam, NL-1018XE Amsterdam, The Netherlands}

\author{R. Ovsyannikov}
\affiliation{Helmholtz Zentrum Berlin GmbH, Albert-Einstein-Strasse 15, 12489 Berlin, Germany}

\author{J. Fink}
\affiliation{Helmholtz Zentrum Berlin GmbH, Albert-Einstein-Strasse 15, 12489 Berlin, Germany}
\affiliation{Leibniz Institute of Solid and Materials Research Dresden, P.O. Box 270116, D-01171, Dresden, Germany}

\author{H. A. D\"urr}
\affiliation{Helmholtz Zentrum Berlin GmbH, Albert-Einstein-Strasse 15, 12489 Berlin, Germany}

\author{A. Gloskovskii}
\affiliation{Institut f\"ur Anorganische Chemie und Analytische Chemie, Johannes Gutenberg-Universit\"at, 55099 Mainz,
Germany}

\author{H. S. Jeevan}
\affiliation{I. Physikalisches Institut, Georg-August-Universit\"at, G\"ottingen 37077, Germany}

\author{P. Gegenwart}
\affiliation{I. Physikalisches Institut, Georg-August-Universit\"at, G\"ottingen 37077, Germany}

\author{A. Erb}
\affiliation{Walther-Meissner-Institut, Walther-Mei\ss ner Strasse 8, 85748 Garching, Germany}

\author{L. Patthey}
\affiliation{Paul Scherrer Institut, Swiss Light Source, 5232 Villigen, Switzerland}

\author{M. Shi}
\affiliation{Paul Scherrer Institut, Swiss Light Source, 5232 Villigen, Switzerland}

\author{R. Follath}
\affiliation{Helmholtz Zentrum Berlin, Albert-Einstein-Strasse 15, 12489 Berlin, Germany}

\author{A. Varykhalov}
\affiliation{Helmholtz Zentrum Berlin, Albert-Einstein-Strasse 15, 12489 Berlin, Germany}

\author{M. S. Golden}
\affiliation{Van der Waals-Zeeman Institute, University of Amsterdam, NL-1018XE
Amsterdam, The Netherlands}

\date{\today}

\begin{abstract}
Using angle resolved photoemission it is shown that the low lying electronic states of the iron pnictide parent compound EuFe$_2$As$_2$ are strongly modified in the magnetically ordered, low temperature, orthorhombic state compared to the tetragonal, paramagnetic case above the spin density wave transition temperature. Back-folded bands, reflected in the orthorhombic/ anti-ferromagnetic Brillouin zone boundary hybridize strongly with the non-folded states, leading to the opening of energy gaps. As a direct consequence, the large Fermi surfaces of the tetragonal phase fragment, the low temperature Fermi surface being comprised of small droplets, built up of electron and hole-like sections. These high resolution ARPES data are therefore in keeping with quantum oscillation and optical data from other undoped pnictide parent compounds.
\end{abstract}

%Electronic structure, Manganites, Photoemission
%\keywords{Suggested keywords}%Use showkeys class option if keyword
                              %display desired

\maketitle 

%introduction:

The electronic structure and properties of the newly discovered iron-pnictide superconductors \cite{PNoriginal} are a focus of much research.
A central theme is the understanding of the origins of the magnetic ordering and its possible interplay with superconductivity.
The characteristics of the low-temperature, orthorhombic, anti-ferromagnetic (AFM) phase have been experimentally determined, for example for the undoped \textit{122}-parent compounds, $M$Fe$_2$As$_2$ (with $M=$ Ca, Sr, Ba, Eu\ldots).
A common proposition is that the magnetism is of an itinerant, spin density wave (SDW) type, connected to nesting of the warped cylindrical Fermi surfaces (FS's) centered at the $(0,0)$ ($\Gamma$) and $(\pi,\pi)$ points (X) of the tetragonal Brillouin zone (BZ) \cite{SDW}.
Such an SDW would have dramatic consequences for the band structure and Fermi surface, leading to reconstruction, the opening of gaps and major Fermi surface depletion.
Recently, quantum oscillation (QO) experiments have presented evidence for Fermi surfaces comprising small pockets - due to the effects of the SDW order - in SrFe$_2$As$_2$ \cite{Sebastian} and BaFe$_2$As$_2$ \cite{Analytis}.
Both studies find the existence of three distinct orbits, with FS areas of only 0.3\%, 0.6\% and 1.5\% of the tetragonal BZ (compared to a total FS area larger than 20\% in the tetragonal phase).
In addition, the opening of gaps as well as a dramatic reduction of the free charge carrier density upon entering the orthorhombic AFM state has been concluded from optical conductivity measurements \cite{Optics_Hu, Wu}.
Given that a FS-nesting-driven SDW is rooted in k-space, it is of great importance whether the SDW `fingerprints' of reconstruction, gapping and FS depletion really take place in the $E(k)$-space hosting the electronic states of these materials.
Angle-resolved photoemission (ARPES) is a powerful probe of such issues, and consequently there have been numerous studies of the parent compounds of the pnictide superconductors using ARPES.
Early studies were unable to detect significant changes below the magnetic ordering temperature \cite{Liu_Kam_PRL101,Us_1stPRB}.
More recently, ARPES studies have shown the existence of small, additional FS pockets (either hole or electron-like) around the $(0,0)$ or $(\pi,\pi)$ point \cite{EvPnic,Kaminski_condmat,Hsieh,Richard_arXiv:0909.0574}, and a detailed temperature-dependent study is reported in \cite{Yi}.
Despite this progress, it could be argued that there are no studies that cover the states around $\Gamma$ and $X$ on an equal footing, avoiding matrix-element-induced extinction of important states, and there are, in any case, no studies reported of the europium parent compound. 
On top of this, the issues of the dimensionality of the electronic states \cite{Kam_dim,Mannella,Brouet,Us_dim} and the nature of the cleavage surface of the $M$Fe$_2$As$_2$ systems \cite{Freek_Surface,Gao} are both matters of quite some discussion.

\begin{figure} [!t]
\begin{center}
\includegraphics[width=1.0\columnwidth]{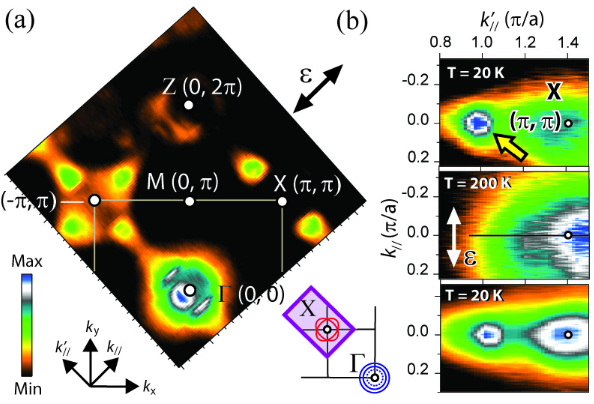}
\caption{Temperature dependent Fermi surface (FS) topology of Eu122. (\textbf{a}) Measured FS (E$_{F}\pm10$~meV, h$\nu=84$~eV) of Eu122 at $T=10$~K. The light polarization $\epsilon$ is indicated. The inset shows a schematic high temperature FS in a two-dimensional projection covering a quarter of the tetragonal Brillouin zone. (\textbf{b}) FS maps (h$\nu=75$~eV) around the $(\pi,\pi)$-point in $k$-space (purple box in the inset of [a]) at different temperatures. The polarization vector ($\epsilon$) is parallel to the analyzer slit. The top-most 20~K FS map is taken on a fresh cleavage surface, the bottom one after cycling the sample temperature up to 300~K \cite{kf_comment}.}
\label{fig:pic1}
\end{center}
\end{figure}

In this paper we present the first photoemission study on the pnictide parent compound Eu122, showing data of a clarity that makes use of second differentials unnecessary. The local iron moment, $\approx1\mu_B$ \cite{Xiao}, for Eu122 is similar to the other $M$122 compounds, yet in Eu122 the Eu moments also order at low temperatures, having a local moment of nearly $7\mu_B$. Our low temperature ARPES data however do closely resemble those reported for other parent compounds \cite{Hsieh, Yi}, indicating that the ordering of the Eu moments is of little influence on the ordering of the iron moments.

Exploiting matrix elements and polarisation-dependent selection rules, we uncover hybridization effects between back-folded hole and electron-like bands both at the $(\pi,\pi)$ and $(0,0)$ regions of k-space and provide clear evidence of the resulting energy gaps. As a result, the large FS's of the paramagnetic phase break up into very small hole and electron pockets we dub `droplets' on entering the magnetically ordered state. The exact nature of the AFM phase (localized vs. itinerant) is still subject of debate \cite{Han, Eu_loc}. We note that our ARPES measurements cannot a priori distinguish between localized versus itinerant magnetic order. Even more so, the orthorhombic structural transition should also give rise to an electronic reconstruction. Since the orthorhombic and magnetic transitions coincide in the parent compounds, it is virtually impossible to disentangle the effect of the combined electronic reconstructions. Regardless of its origin, the back-folding, hybridization and gapping we observe in our ARPES data are consistent with the QO and optical conductivity results on related compounds.

%Experimental
Temperature dependent ARPES was performed at the UE112-PGM beamline at BESSYII, coupled to a \textit{Scienta} SES100 analyzer with a total energy resolution of 27~meV (at $T=20$~K, h$\nu=75$~eV) and a resolution of better than $0.02$~$\pi/a$ along the slit. The analyzer slit was perpendicular to the scattering plane of the incoming light and the outgoing photoelectrons and the polarization vector of the light was perpendicular to this plane. Additional, detailed FS mapping experiments were performed at the SIS-HRES beamline of the Swiss Light Source, using a \textit{Scienta} R4000 analyzer with a total energy resolution of 16~meV (at $T=10$~K, h$\nu=84$~eV) and a momentum resolution better than $0.05$~$\pi/a$ along the slit. FS maps were collected with an angular step of 0.5$^{\circ}$, scanning in the direction perpendicular to the analyzer slit (denoted $k'_{\parallel}$ forthwith), which is comparable to the resolution along the analyzer slit. The small beam spot-size ($\approx 100$~$\mu$m) assured a minimal influence of surface roughness on the ARPES experiments and aided in the achievement of very high effective contrast in $k$-space. During the SLS experiments the analyzer slit and the polarization vector of the synchrotron radiation were parallel (p-polarized) to the scattering plane formed by the incoming light and the outgoing photoelectrons.

Single crystals of Eu122 were grown using the Bidgeman and Sn-flux method at
Goettingen University \cite{Gegenwart2}. Their magnetic transition temperature was determined to be about 190~K, with an additional AFM ordering of the Eu moments around 18~K \cite{Gegenwart}. Crystal surfaces were cleaved at $T<50$~K in a vacuum better than $1\times10^{-10}$~mbar. The quality of the cleavage surfaces was checked using low energy electron diffraction (LEED). All measured cleaves yielded very sharp diffraction patterns, displaying, apart from the expected $1\times 1$ spots, prominent $2\times1$, $1\times2$, and less clear $\sqrt{2}\times\sqrt{2}$ reconstruction spots. These surface reconstructions were not robust against thermal cycling up to room temperature, in keeping with their origins in ordering/disordering processes within the partial Eu layer that comprises the termination of the cleaved crystal \cite{Freek_Surface,Gao}.

\begin{figure} [!b]
\begin{center}
\includegraphics[width=1.0\columnwidth]{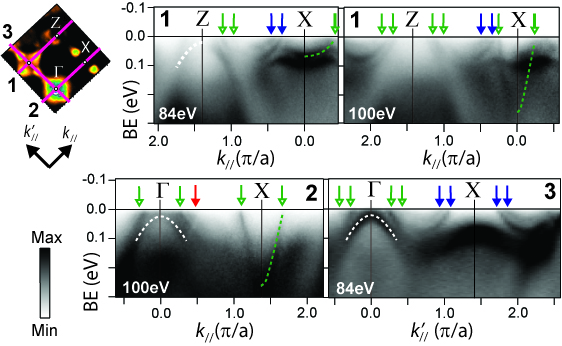}
\caption{Low temperature Fermi surface topology of Eu122 ($T=10$~K, h$\nu=84$ and $100$~eV). The $(E,k)$-maps show raw data taken from the cuts in $k$-space indicated with pink lines in the small FS map. For an explanation for colored the arrows, see text.}
\label{fig:pic2}
\end{center}
\end{figure}

%Body: Results
We start with a discussion of the `as cleaved', $T=10$~K data from EuFe$_2$As$_2$ shown in Fig. \ref{fig:pic1}a. The inset shows a cartoon of the $k$-space region under scrutiny, with a simplified representation of the band theory predictions for the tetragonal phase: three hole pockets around $(0,0)$ (blue circles) and two electron pockets at $(\pi,\pi)$ (red ellipses) \cite{Ma}. The main part of Fig. \ref{fig:pic1}a is a wide k-range, low temperature FS map which obviously exhibits quite a different FS topology than the tetragonal expectation.

The most eye-catching difference is the fact that both the large hole-like FS's at $(0,0)$ and the electron-like FS's at $(\pi,\pi)$ and equivalent points have broken up into a number of clearly disconnected `droplets' \cite{pol_note}. We stress that our data clearly resolves the droplet structure of the FS's at both high symmetry points: there are no large FS's left.
Droplet, or petal-like FS's have been reported in ARPES studies of differing $k$-space locations of the Sr122 (Ref. \cite{Hsieh}), Ba122 (Ref. \cite{Kaminski_condmat,Yi}) and Ca122 (Ref. \cite{Kaminski_PRL}) parent compounds, as well as from the doped superconductor (Ba,K)Fe$_2$As$_2$ (Ref. \cite{EvPnic}). We will zoom in on the $\Gamma$ and $X$ $k$-space regions for Eu122 later in the paper. Before doing this, we deal with the issue of surface reconstructions. 

Seeing as the simplest SDW scenario involves shifted copies of bands appearing at new $k$-space locations compared to the high-$T$ tetragonal phase, it is important to test the impact of (surface) structural reconstructions on the data, particularly as such reconstructions are a common feature of scanning tunneling microscopy data from the $M$Fe$_2$As$_2$ materials \cite{Freek_Surface,PhysicaC_STM_rev}. Given the fact that the extra LEED spots, arising from the ordering of the $M$-cations in the surface layer, disappear irreversibly on warming the (cold cleaved) sample up to room temperature, re-cooling then offers us the test we require.

In Fig. \ref{fig:pic1}b, we show that the FS's around $(\pi,\pi)$ remain unchanged upon raising the temperature from 10~K to 20~K, at which point the Eu moments have become disordered \cite{Eu-comment}, illustrating that the magnetic ordering of the Eu sub-lattice does not have a big effect on the Fe-derived electronic states at the Fermi level as seen with ARPES. The situation is very different at $T=200$~K: upon exiting the AFM phase of the Fe-planes, the droplet FS structure (indicated with an arrow and appearing as a dumb-bell in the measurement geometry used in Fig. \ref{fig:pic1}b due to intensity from a below-$E_F$ band at the X point itself) is no longer observed. At 200~K, the LEED pattern shows only 1x1 tetragonal spots. Upon re-cooling to 20K (bottom panel of Fig. \ref{fig:pic1}b), the extra, $2\times1$ and $\sqrt{2}\times\sqrt{2}$ LEED spots do not re-appear, yet the droplet FS at $(\pi,\pi)$ does, proving that this remarkable FS topology reflects the bulk, magnetic state of the pnictide parent compound, and is definitely not a result of `diffraction replicas' due to surface reconstructions.

\begin{figure} [!t]
\begin{center}
\includegraphics[width=1.0\columnwidth]{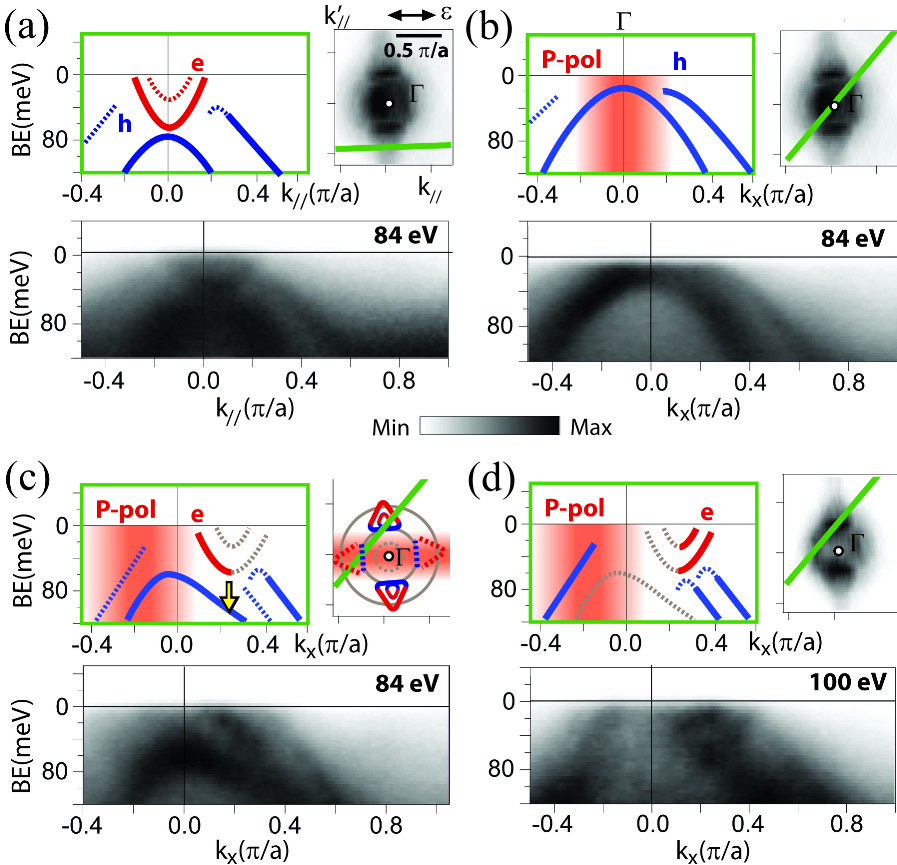}
\caption{Detailed FS topology around the $(0,0)$-points of Eu122 ($T=10$~K, h$\nu=84$ (Z) and $100$~eV ($\Gamma$). For each of (\textbf{a})-(\textbf{d}), the left panels show schematic (top) and measured (bottom) $(E,k)$-images for the green $k$-space cuts shown in the small FS maps (top right panel in each case). In each band structure scheme, electron-pocket bands are labeled `e' and indicated in red, whereas hole-pocket bands get an `h', and are shown in blue. Barely distinguishable features are indicated with dotted lines. Red shading indicates areas in which the symmetry selection rules can lead to suppression of some of the orbital states. The measured FS map in panel (c) has been replaced with a sketch of the low $T$ Fermi surface. The FS predicted for the tetragonal phase is shown in gray.}
\label{fig:pic3}
\end{center}
\end{figure}

Now we are sure that the observed low-$T$ Eu122 FS is sensitive to the bulk magnetism, we return to the band structure in the orthorhombic phase in more detail.  In Fig. \ref{fig:pic2} several $E(k)$-cuts through the $\Gamma$ and Z points, parallel to $\Gamma$$X$ are shown, taken with h$\nu=~$100 and 84~eV.  The data taken with these two photon energies look, in general, very similar - with FS crossings at very similar $k_F$'s, albeit with (at times strongly) varying spectral weight. As these two photon energies correspond to $k_z = 0$ (100~eV) and $\pi$/c (84~eV), this illustrates that the $k_z$-dispersion for this compound is modest \cite{kz_note}. A second point is that the destruction of the large FS's to give droplets is clearly present in Eu122 at both $k_z = 0$ and $k_z = \pi/c$, unlike what has been reported for Ba and Ca122 \cite{Kaminski_condmat}.

We note that the surface quality and favorable experimental conditions mean that a large number of band features can be distinguished in the $E(k)$ images shown in Fig. \ref{fig:pic2}. We deal firstly with a group of features that clearly possess reversed character with respect to the high-$T$, tetragonal situation, being hole-like around X and electron-like around $\Gamma$(Z). These states are marked in Fig. \ref{fig:pic2} using blue and red arrows, for the hole-like and electron-like features, respectively, and originate from the effects of magnetic order in the Fe plane.

Further inspection of Fig. \ref{fig:pic2}, however, shows that there are also ARPES features (marked with green arrows) which seem to match those expected in the paramagnetic, tetragonal state, such as the electron-like feature clearly visible around $(\pi,\pi)$ for h$\nu=$~100eV. This apparent dichotomy - that both reconstructed and unreconstructed bands coexist - has been reported in ARPES investigations of other parent compounds,\cite{Hsieh,Kaminski_condmat} and will be returned to further on. Dealing with these kind of issues one is confronted with the intrinsic complexity of the low lying electronic structure of these multiband materials. Even adopting the simplest picture of only back-folding the tetragonal band structure yields 8 or 10 bands (starting from 4 or 5 FS's at high temperature), and if the crystal contains small-scale magnetic twin domains we average over in the ARPES experiment, then this number doubles. It is therefore imperative to zoom-in and take a closer look at the different high symmetry points in $k$-space.

To start off, we zoom-in on the $\Gamma$-point in Fig. \ref{fig:pic3}. Four $k$-space cuts for the two different photon energies are presented, partnered with a band structure scheme as a guide to the $E(k)$ images. Panel (a) clearly shows the existence of shallow electron-like bands, appearing in the FS map as small, triangular shaped pockets. Cuts through the center of this feature along the $k_x$-direction using h$\nu=84$ and $100$~eV are shown in panels (c) and (d), and reveal that, in fact, it consists of two concentric electron pockets. The cuts shown in Fig. \ref{fig:pic3}(b) illustrate a d-wave like gapping operative around the $(0,0)$ point, with no low energy pockets such as those seen at $k_x\sim0.2\pi/a$ in panels (c) and (d). The concentric electron pockets have Fermi surface areas of about 1\% and 0.3\% of the tetragonal BZ. Such small FS areas match those observed in the QO studies of Sr122 and Ba122, although we note that the details of the band structure of the different pnictide parent compounds will depend somewhat on the identity of the $M$-cation.

The important point here is the fact that we are able to detect the disintegration of the large $\Gamma$-centered FS's into very small droplets. As can be seen in the $E(k)$ images, the two different photon energies either more clearly resolve the left (84~eV) or the right side (100~eV) of the small pockets, and this matrix element difference is, in fact, a signal of different orbital character for either side of the pocket. A natural explanation for this would be that these multiple, orbitally-split pockets consist of back-folded and hybridized hole-like and electron-like bands which started life as the $\Gamma$ and X-point band structure features in the tetragonal phase, respectively. 

As the $E(k)$-sketches indicate, the data still contains remnants resembling the tetragonal bands. However, these interact with the back-folded bands such that the former do not cross the Fermi energy, and thus do not give rise to large FS's around $\Gamma$. In Fig. \ref{fig:pic3}(c) one can also see a further case of interaction between a folded and non-folded band in that the blue hole-like band - which is well below the Fermi level - is not symmetric around $k_x=$~0 but is further pushed down by the presence of the lower binding energy back-folded bands on the right. These data, then, do not support a picture of selective modification of bands from the tetragonal band structure,\cite{Hsieh} in which only some bands interact with the back-folded structures, a point we will return to in the following discussion of the X-point zoom-in.

\begin{figure} [!t]
\begin{center}
\includegraphics[width=1.0\columnwidth]{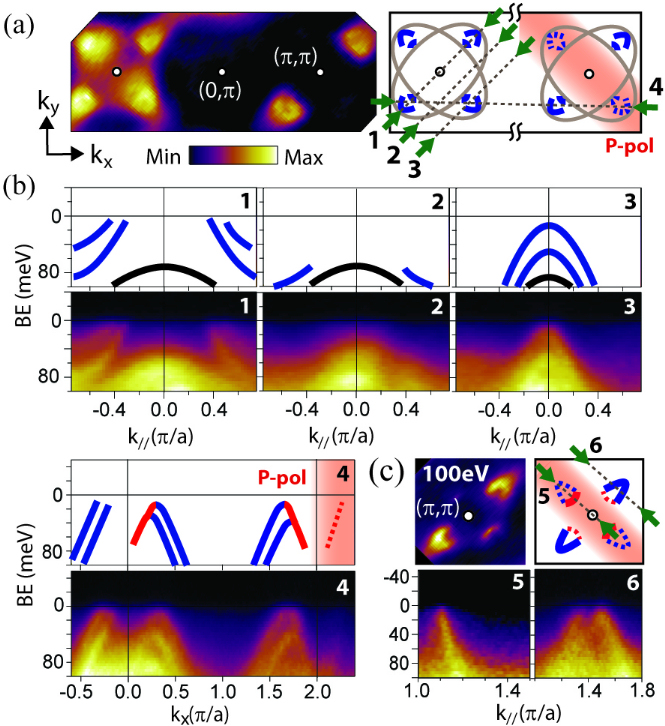}
\caption{Detailed X-point Fermi surface topology of Eu122 ($T=10$~K, h$\nu=84$~eV). (\textbf{a}) FS map covering the $(\pi,\pi)$-$(-\pi,\pi)$-$(\pi,\pi)$ region of $k$-space (left) with a 'guide' sketch (right), including the high temperature Fermi surfaces in gray. (\textbf{b}) $(E,k)$-maps taken from the cuts in $k$-space indicated in (a) partnered by sketches in which electron-like bands are indicated in red and hole-like in blue. (\textbf{c}) X-point FS map, h$\nu=100$~eV, including a sketch showing the location of $k$-cuts 5 and 6. As in Fig. 3, the red shading marks the effect of the symmetry selection rules for certain orbital states.}
\label{fig:pic4}
\end{center}
\end{figure} 

We now turn to the states around the X or $(\pi,\pi)$ point, which - in general - have received less attention in ARPES papers on the parent compounds than have those around $\Gamma$. 
In Fig. \ref{fig:pic4}(a) and (b), data recorded with 84eV photons are presented. Cut 1 in panel (b) clearly shows two pairs of hole-like features crossing $E_F$, giving rise to the two pairs of blue features shown along cut 1 in the main FS sketch [panel (a)]. This is in contrast to cut 2, which reveals that the hole-like bands do not approach $E_F$: their spectral weight is suppressed up to a binding energy of about 70~meV, which, as also noted in Ref. \cite{Hsieh}, corresponds rather well to the gap observed in optical conductivity measurements (with $2\Delta\approx90$~meV for Eu122 \cite{Wu}). Thus, cut 2 of Fig. \ref{fig:pic4} attests to the opening of a gap, leading to the break-up of the large elliptical X-point FS's into significantly smaller structures.  

The data of cut 3 take us a step further: here the cut just grazes the end of the longer hole-like arc, and if we take this result at face value the droplet-like FS would, in fact, have to be `open'. Thus, for the $k_z$ value close to $\pi$/c (i.e., that relevant for h$\nu=84$~eV) if we were conducting a quantum oscillation experiment, there would be no closed FS orbits measurable in the $k_{x,y}$ plane near $(\pi,\pi)$.
The $E(k)$-image along $k$-space cut 4 clearly shows that the hole-like bands bend back into an electron-like feature around X that has its band bottom at a binding energy $BE \approx 100$~meV. This delivers the clue that the crescent-shaped, gapped, FS segments around the X-point in the AFM orthorhombic phase arise from hybridization of the X-centered electron pockets with the back-folded hole-like bands from the $\Gamma$ point.

We now switch in Fig. \ref{fig:pic4}(c) to data recorded with h$\nu=100$~eV (for which $k_z\approx0$). Comparing the FS maps of panels (a) and (c) and the data of cuts 1 and 5, we note that for 100~eV photons a clear electron-like feature is observed, located at the opening of the hole-like arc seen also for 100~eV photons in cut 6. These electron-like sections can be seen to act as a `gate', closing off the arc into a continuous, droplet-like FS.

Thus, these data from the X-point can support two scenarios: i) firstly, if the lack of hole-like bands in cut 1 is taken to be an initial state effect (i.e., is not simply due to a low matrix element), then we see a $k_z$-dependence in which for $k_z=$~0 (h$\nu=100$~eV) there are small FS droplets around the $(\pi,\pi)$ points, whereas for $k_z=$~$\pi$/c (h$\nu=84$~eV) the hole-like arcs remain open and thus no closed FS exists near X for this $k_z$ value. 
ii) the alternative is to posture that there are electron-like `gates' closing the hole-like arcs for all $k_z$ values (i.e., the droplets are, in fact thin, roughly triangular prisms with their long axis along $k_z$), but that they carry no spectral weight for h$\nu=84$~eV. 

We believe scenario (ii) to be more likely, but irrespective of the $k_z$ dependence, the large FS pockets present in the paramagnetic phase are broken up into much smaller structures on entering the AFM phase, either with or without continuity along $k_z$. 

This first ARPES study of the EuFe$_2$As$_2$ parent compound of the pnictide high $T_c$ superconductors has yielded data of high quality, with excellent `contrast' in $k$-space. This enables a detailed discussion of all Fermi surfaces near both the $(0,0)$ and $(\pi,\pi)$ points on an equal footing. Thus, bringing together the conclusions from both the $\Gamma$ and X-point data, we can state the following by way of summary:
\begin{enumerate}
\item
On cooling well below the ordering temperature of the iron planes, the large FS's (reversibly) break-up into small droplet-like structures, supporting only a fraction of the original, high temperature FS areas.  
\item
The fact that both electron-like features are found at $(0,0)$ and hole-like at $(\pi,\pi)$ suggests that back-folding of the high-T band structure takes place via reflection in the new Brillouin zone boundaries.
\item 
The back-folding picture is strongly supported by clear evidence for hybridization between the back-folded and non-folded bands, resulting in shifting of bands and the opening of d-wave-like energy gaps of order 50-70meV. 
\item
Comparison of data probing different locations along the $k_z$ axis suggest that either the small FS structures near the $(\pi, \pi)$ point might indeed be droplets --without continuity along $k_z$-- or they take the form of prisms stretching along $k_z$.  
\item
Finally, the ordering of the Eu moments below ca. 20K does not have a noticeable impact on the ARPES data. 
\end{enumerate}

Taking points $2)-4)$ together, we can close by stating that these ARPES data are in keeping with the conclusions from recent quantum oscillation and optical data on other $M$Fe$_2$As$_2$ parent compounds \cite{Sebastian,Analytis,Optics_Hu,Wu}, that offer support for a large suppression of the number of near $E_F$ electronic states in the low temperature magnetically ordered orthorhombic phase.

%Acknowledgement
We acknowledge the IFW Dresden for the provision of their SES100 spectrometer via the Helmholtz Zentrum Berlin, and Huib Luigjes for expert technical support. This work is part of the research programme of the `Stichting voor Fundamenteel Onderzoek der Materie' (FOM). We also acknowledge funding from the EU (via I3 contract RII3-CT-2004-506008).


\begin{thebibliography}{}
\bibitem{pol_note} At the $(\pi,\pi)$ point in Fig. 1a, only two of the four droplet FS's are visible, due to the strong polarisation dependence of the photoemission matrix elements, linked to the presence of a mirror plane in the Fe-plane and the p-polarised synchrotron radiation used. For $k'_{\parallel}$ values well away from zero, this selection rule no longer operates and all four droplet FS's are seen, such as at $(-\pi,\pi)$.
\bibitem{kz_note} The assignment of h$\nu=84$~eV as $k_z=\pi$/c and h$\nu=100$~eV as $k_z=0$ is strictly speaking only valid at the $\Gamma$ point: the $k_z$ vectors at the X-point are about 15\% of the reciprocal c-axis lattice vector smaller.
\bibitem{kf_comment} Between the two data sets recorded at $T=20$~K in panel (c) of Fig. \ref{fig:pic1}, several warm up- cool down cycles were performed, with slightly different spectral weight distribution and $k_F$ loci for the droplet-like FS seen between the two spectra. The mechanism leading to these small changes is not clear, but the important, quantitative point to note is the dis- and re-appearance of the small FS droplets above and below the iron magnetic ordering temperature.
\bibitem{Eu-comment} We carried out further checks whether the ordering of the Eu moments had any influence on the electronic structure of Eu122 as probed with ARPES, but no difference was found for several different $k$-cuts taken at 10K and well above 20K.
\end{thebibliography}
\end{document}